\documentclass[a4paper,fleqn,usenatbib]{mnras}

\usepackage{mathptmx}
\usepackage[T1]{fontenc}
\usepackage{ae,aecompl}

\usepackage{graphicx}	% Including figure files
\usepackage{amsmath}	% Advanced maths commands
\usepackage{amssymb}	% Extra maths symbols

\newcommand{\pcm}{\,cm$^{-2}$}	% per cm-squared

\newcommand{\NSCool}{\textit{NSCool}}

\newcommand{\Ts}{\ensuremath{T_{\rm s}}}
\newcommand{\Tb}{\ensuremath{T_{\rm b}}}
\newcommand{\Msun}{\ensuremath{M_{\rm Sun}}}
\newcommand{\rhob}{\ensuremath{\rho_{\rm b}}}

\newcommand{\Tso}{\ensuremath{T_{\rm s,old}}}
\newcommand{\yh}{\ensuremath{y_{\text{H}}}}
\title[The effect of DNB on NS crust cooling]{The effect of diffusive nuclear burning in neutron star envelopes on cooling in accreting systems}

\author[M.J.P. Wijngaarden et al.]{
M.J.P. Wijngaarden,$^{1}$\thanks{E-mail: M.J.P.Wijngaarden@soton.ac.uk}
Wynn C.G. Ho$^{1,2}$,
Philip Chang$^{3}$,
Dany Page$^{4}$,
\newauthor
~Rudy Wijnands$^{5}$,
Laura S. Ootes$^{5}$,
Andrew Cumming$^{6}$,
Nathalie Degenaar$^{5}$,
\newauthor
~Mikhail Beznogov$^{4}$
\\
$^{1}$Mathematical Sciences and STAG Research Centre, University of Southampton, SO17 1BJ, Southampton, UK\\
$^{2}$Department of Physics and Astronomy, Haverford College, 370 Lancaster Avenue, Haverford, PA, 19041, USA\\
$^{3}$Department of Physics, University of Wisconsin-Milwaukee, 1900 E. Kenwood Blvd., Milwaukee, Wisconsin 53211, USA\\
$^{4}$Instituto de Astronom\'ia, Universidad Nacional Aut\'onoma de M\'exico, Mexico City, CDMX 04510, Mexico\\
$^{5}$Anton Pannekoek Institute for Astronomy, University of Amsterdam, Postbus 94249, 1090 GE Amsterdam, The Netherlands\\
$^{6}$Department of Physics, McGill Space Institute, McGill University, 3550 rue University, Montreal, QC H3A 2T8, Canada
}

\date{Accepted 2020 February 26. in original form 2020 January 22}

\pubyear{2019}

\begin{document}
\label{firstpage}
\pagerange{\pageref{firstpage}--\pageref{lastpage}}
\maketitle

% Abstract of the paper
\begin{abstract}
Valuable information about the neutron star interior can be obtained by comparing observations of thermal radiation from a cooling neutron star crust with theoretical models. Nuclear burning of lighter elements that diffuse to deeper layers of the envelope can alter the relation between surface and interior temperatures and can change the chemical composition over time. We calculate new temperature relations and consider two effects of diffusive nuclear burning (DNB) for H-C envelopes. First, we consider the effect of a changing envelope composition and find that hydrogen is consumed on short timescales and our temperature evolution simulations correspond to those of a hydrogen-poor envelope within $\sim$100 days. The transition from a hydrogen-rich to a hydrogen-poor envelope is potentially observable in accreting NS systems as an additional initial decline in surface temperature at early times after the outburst. Second, we find that DNB can produce a non-negligible heat flux, such that the total luminosity can be dominated by DNB in the envelope rather than heat from the deep interior. However, without continual accretion, heating by DNB in H-C envelopes is only relevant for $<$1-80 days after the end of an accretion outburst, as the amount of light elements is rapidly depleted. Comparison to crust cooling data shows that DNB does not remove the need for an additional shallow heating source. We conclude that solving the time-dependent equations of the burning region in the envelope self-consistently in thermal evolution models instead of using static temperature relations would be valuable in future cooling studies. 
\end{abstract}

% Select between one and six entries from the list of approved keywords.
% Don't make up new ones.

%%%%%%%%%%%%%%%%%%%%%%%%%%%%%%%%%%%%%%%%%%%%%%%%%%

%%%%%%%%%%%%%%%%% BODY OF PAPER %%%%%%%%%%%%%%%%%%

\section{Introduction}
\label{sec:introduction}

% Neutron star cooling/temperature evolution helps understand interiors
% Cooling long term + accreting systems
Neutron stars (NSs) are an excellent laboratory for investigating extreme physics over a large range of densities, including supranuclear densities in their cores. One method to probe NS interiors is comparing theoretical models with observed temperatures of cooling NSs. If left undisturbed, isolated NSs cool over time after they are born hot in supernova explosions (see, e.g., \citealt{2015SSRv..191..239P}). NSs in binary systems can occasionally undergo accretion outbursts, when new material is accreted from the companion star onto the NS surface. During an accretion outburst, the NS crust can be heated out of thermal equilibrium with the core as a sequence of non-equilibrium nuclear reactions in the crust, collectively known as {\em deep crustal heating}, are triggered when the underlying material is compressed by newly accreted matter \citep{1990A&A...227..431H,1998ApJ...504L..95B}. After the accretion outburst, the NS crust cools back to thermal equilibrium with the core on time scales of years (see, e.g., \citealt{2017JApA...38...49W} for a review). This means that the cooling phase can be covered with multiple observations during which the surface temperature can be measured to constrain theoretical cooling curves, which is a useful opportunity to gain insight in the properties and physics of the NS crust and core (e.g., \citealt{2013PhRvL.111x1102P,2017PhRvC..95b5806C,2018PhRvL.120r2701B}).

For several crust cooling sources, the high observed surface temperatures $\lesssim$ 100 days after the accretion outburst can not be explained using the standard deep crustal heating model but require the presence of an additional shallow heating source (at densities $\rho < 10^{11}$ g cm$^{-3}$) during outburst. Typically, during the outbursts, an amount of $\sim$1-2 MeV per accreted nucleon  of shallow heating is needed to explain the observed cooling curves after the outbursts are over (e.g., \citealt{2009ApJ...698.1020B,2014ApJ...791...47D,2016ApJ...833..186M}), but for one source (MAXI J0556-332) as much as $\sim$17 MeV nucleon$^{-1}$ was needed \citep{2014ApJ...795..131H,2015ApJ...809L..31D,2017ApJ...851L..28P}. This means that the amount of shallow heating can be larger than that due to deep crustal heating, which releases $\sim$2 MeV nucleon$^{-1}$. One of the outstanding problems in cooling studies is the unknown physical mechanism of this shallow heat (see the discussion in \citealt{2015ApJ...809L..31D}). Thus, it is useful to explore additional heating mechanisms that are currently not accounted for in the present crustal heating models.

% Uncertainty in cooling studies due to unknown envelope composition (constrain from x-ray burst?)
One of the challenges in understanding crust cooling and shallow heating is obtaining the interior temperature (at the bottom of the envelope) from observations of the surface emission. The relation between surface temperature and interior temperature is set by the heat conducting properties of the thin outer envelope ($\rho < \rhob = 10^{8} - 10^{10}$ g cm$^{-3}$), which is highly sensitive to chemical composition \citep{1983ApJ...272..286G,1997A&A...323..415P,2002ApJ...574..920B}. NS cooling codes typically calculate the interior structure and thermal evolution from the centre of the star out to the bottom of the envelope (at $\rhob$). Then, analytic fits of the temperature relations between the NS surface temperature ($\Ts$)~and the temperature at the bottom of the envelope, $\Tb$ [$\equiv T(\rhob)$], are used as a boundary condition. Analytic fits to numerical envelope calculations for different static envelope compositions and envelope sizes have been calculated for one- \citep{1983ApJ...272..286G} and two or more chemical component models (see, e.g., \citealt{1997A&A...323..415P,2003ApJ...594..404P,2016MNRAS.459.1569B,2019MNRAS.484..974W}).

% DNB can change the envelope composition
Diffusive nuclear burning (DNB) of hydrogen and helium can affect the envelope \Ts-\Tb~relations and can alter the envelope composition over time. DNB occurs when lighter elements diffuse to depths where the temperature and density are large enough to ignite nuclear burning of these elements. This process has been studied for H-C \citep{2003ApJ...585..464C,2004ApJ...605..830C} and He-C envelopes \citep{2010ApJ...723..719C}. In addition, \cite{2019MNRAS.484..974W} calculated H-He and He-C temperature relations including the effect of DNB and investigated how DNB can alter the envelope composition over time for cooling isolated neutron stars.

Here we obtain analytic temperature relations for H-C envelopes and explore the effect of DNB on the observed cooling of NSs after an accretion outburst. The temperatures in the envelope after accretion outbursts are sufficiently large such that the heat deposited in the envelope by DNB may not be negligible, as was assumed in previous works. We investigate whether the common assumption in cooling studies that no energy is generated in the envelope during quiescence and that its composition does not change significantly during the cooling time is valid when DNB is taken into account. We compare the luminosity generated by DNB in the envelope when the NS is cooling to the interior luminosity, to explore the role of DNB as an additional heating mechanism. We consider how the envelope composition changes over time due to the consumption of lighter elements, which by itself can alter the observed cooling curves as the heat conduction properties of the envelope are sensitive to its chemical composition. 

\section{H-C envelopes}

\subsection{\Ts - \Tb~relations}
\label{sec:tstb}

\begin{figure}
	% To include a figure from a file named example.*
	% Allowable file formats are eps or ps if compiling using latex
	% or pdf, png, jpg if compiling using pdflatex
	\includegraphics[width=\linewidth]{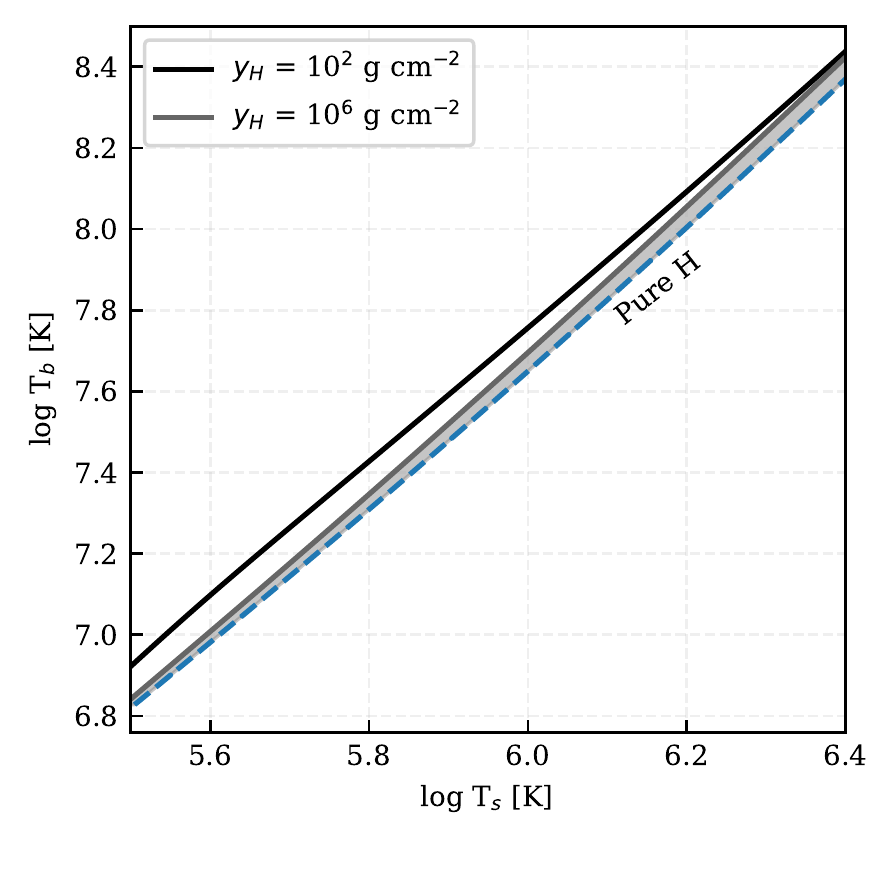}	
	\vspace{-20pt}
	\caption{The $\Ts$-$\Tb$ relation for the H-C envelope with $\rhob$~=~10$^{10}$~g~cm$^{-3}$. The solid line corresponds to both the model with and without DNB, as they are indistinguishable. The blue dashed line corresponds to an hydrogen column of 10$^{10}$~g~cm$^{-2}$ when DNB is not included. The grey shaded region corresponds to the grey region in Figure \ref{fig:HCtransition} and shows the region in the $\Ts-\Tb$ relations that is excluded by taking DNB into account as the build-up of those hydrogen columns are prevented by nuclear burning. }
	\label{fig:HC_TsTb}
\end{figure}

We calculate new temperature relations that include diffusive nuclear burning for a H-C envelope with $\rhob$ = 10$^{10}$ g~cm$^{-3}$. These temperature relations were calculated by computing a grid of envelope models for different combinations of surface temperatures and hydrogen columns and fitting an analytic relation to these results. More details on this approach can be found in Appendix \ref{sec:appendix_tbts} and \cite{2019MNRAS.484..974W}. In Figure \ref{fig:HC_TsTb}, we show the resulting relation between surface temperature and bottom boundary temperature for varying hydrogen column sizes. For a H-C mixture, increasing the hydrogen column leads to a better heat-conducting envelope (i.e., the same bottom boundary temperature corresponds to a larger surface temperature). Note that increasing the hydrogen column in a H-He envelope has the opposite effect (see, e.g., \citealt{2016MNRAS.459.1569B}). Analytic fits for the \Ts-\Tb~relations as a function of hydrogen column size (\yh) and scalable by surface gravity are given in Appendix \ref{sec:appendix_tbts}.

\begin{figure}
	% To include a figure from a file named example.*
	% Allowable file formats are eps or ps if compiling using latex
	% or pdf, png, jpg if compiling using pdflatex
	\includegraphics[width=\linewidth]{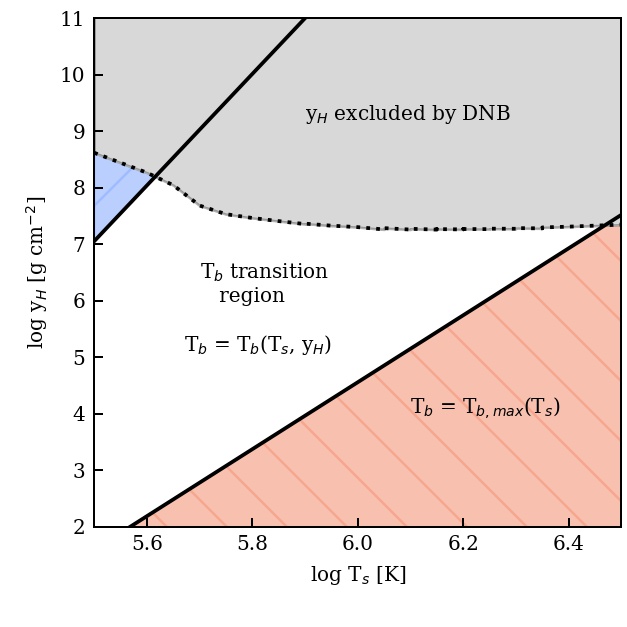}	
	\vspace{-20pt}
	\caption{$\Ts$-$\yh$ parameter space for the H-C envelope model with $\rho_b$ = 10$^{10}$ g cm$^{-3}$. The solid lines enclose the region where the boundary temperature is sensitive to the hydrogen column density, i.e. $\Tb$~=~$\Tb(\Ts,\yh)$. In this region, for a given \Ts~and increasing \yh, the boundary temperature changes from that corresponding to a pure carbon envelope (red), where $\Tb$~=~$T_{\text{b,max}}(\Ts)$, to that of a pure hydrogen envelope (blue), where $\Tb$~=~$T_{\text{b,min}}(\Ts)$. Further changes in the hydrogen column outside of the transition region have negligible effect on the $\Tb$-$\Ts$ relations. The grey shaded region shows the parameter space that is excluded due to DNB (see text). Note that at large $\yh$, some excluded hydrogen columns overlap with the transition region. This means that the range in $\Tb$ is smaller when DNB is taken into account. }
	\label{fig:HCtransition}
\end{figure}

The \Ts-\Tb~relation is highly sensitive to the composition in the \textit{sensitivity strip} of the envelope (where the dominant opacity changes between radiative and conductive). Thus when the change in hydrogen column size does not affect the composition in the sensitivity strip, the effect on the \Ts-\Tb~relation is negligible. This is shown in Figure \ref{fig:HCtransition}, where we explore the parameter space of \Ts~and \yh~ and show for which values the temperature relations are sensitive to \Ts~and \yh. The red and blue regions in Figure \ref{fig:HCtransition} show for which hydrogen column sizes the temperature relations are insensitive to further changes in column size and correspond to the maximum and minimum \Tb~for a given \Ts, respectively.

Including DNB leads to a clear upper limit on the hydrogen column size, as the build-up of larger columns is prevented by rapid hydrogen burning. Similar to what was found for H-He envelopes, we find that hydrogen columns $> 10^{7}$ g \pcm~ cannot be sustained when DNB is taken into account. Note that for most temperatures, this means that a large portion of hydrogen columns located in the transition region are unphysical, leading to a smaller range in allowed boundary temperatures for a given surface temperature. The boundary temperatures corresponding to the excluded hydrogen columns (see Figure \ref{fig:HCtransition}) are highlighted in grey in Figure \ref{fig:HC_TsTb}. As shown by Figure \ref{fig:HC_TsTb} and \ref{fig:HCtransition}, the main effect of DNB for the temperature relations is not the alteration of the composition profile in the sensitivity strip, but the introduction of naturally excluded hydrogen column sizes.

\section{DNB luminosity}
\begin{figure*}
	% To include a figure from a file named example.*
	% Allowable file formats are eps or ps if compiling using latex
	% or pdf, png, jpg if compiling using pdflatex
	\includegraphics[width=\linewidth]{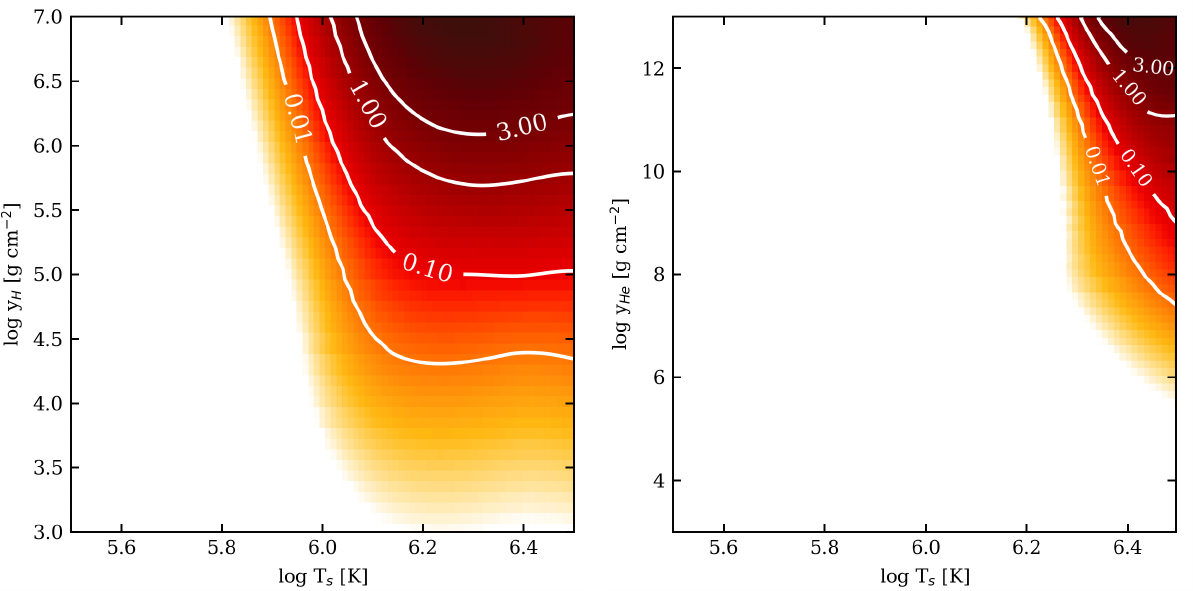}	
	\vspace{-10pt}
	\caption{The explored \Ts~-~$y_{\rm light}$ parameter space for H-C envelopes (left) and He-C envelopes (right). The color is set by the ratio between the DNB luminosity and blackbody luminosity ($L_{\text{DNB}}/L_{\text{int}}$, where $L_{\text{int}} = L_{\text{s}}$ = $4\pi R^2 \sigma \Ts^4$). The colored region shows where the luminosity due to diffusive nuclear burning in the envelope becomes relevant compared to the blackbody luminosity for varying surface temperatures and light element columns. The contour line corresponding to unity indicates where $L_{\text{DNB}}$ becomes larger than $L_{\text{int}}$. The figure parameter ranges are chosen to match relevant values of surface temperatures and light element column densities in crust cooling studies.}
	\label{fig:lums}
\end{figure*}

We compare the total DNB energy deposited in the envelope to the luminosity flowing from the interior (i.e., from any layer deeper than the envelope). The energy generated by DNB is converted to a total DNB luminosity by integrating the local energy generation rate over the mass in the envelope:

\begin{align}
L_{\text{DNB}} = \int \epsilon~\text{d}m = \int \epsilon~4 \pi R^{2} \text{d}y ~{\rm erg~s^{-1}} ,
\end{align} 

\noindent
where $M$ is the total mass of the star, and the mass interval $dm$ is converted to column density interval $dy$ following \cite{1983ApJ...272..286G}.  The local energy generation rate is calculated as 

\begin{align}
\epsilon = \sum \frac{Q_i r_i}{\rho}~{\rm erg~g^{-1}~s^{-1}},
\end{align} 

\noindent
where $Q_i$ is the energy released by a reaction of type $i$, $r_i$ is the reaction rate of reaction type $i$, and $\rho$ is the local density. The luminosity flowing from the interior through envelope, is typically assumed to be constant, such that the luminosity at the surface is $L_{\text{int}}$~=~$L_{\text{s}}$ = $4 \pi R^2 \sigma \Ts^4$ which can be used as a boundary condition in thermal evolution codes (see, e.g., \citealt{2008LRR....11...10C,2009ApJ...698.1020B,2013PhRvL.111x1102P,2015SSRv..191..239P}). 

\begin{figure}
	% To include a figure from a file named example.*
	% Allowable file formats are eps or ps if compiling using latex
	% or pdf, png, jpg if compiling using pdflatex
	%	 Canonical_temp_Teff_tburst_1yr_mdot_1e17.dat
	\includegraphics[width=\linewidth]{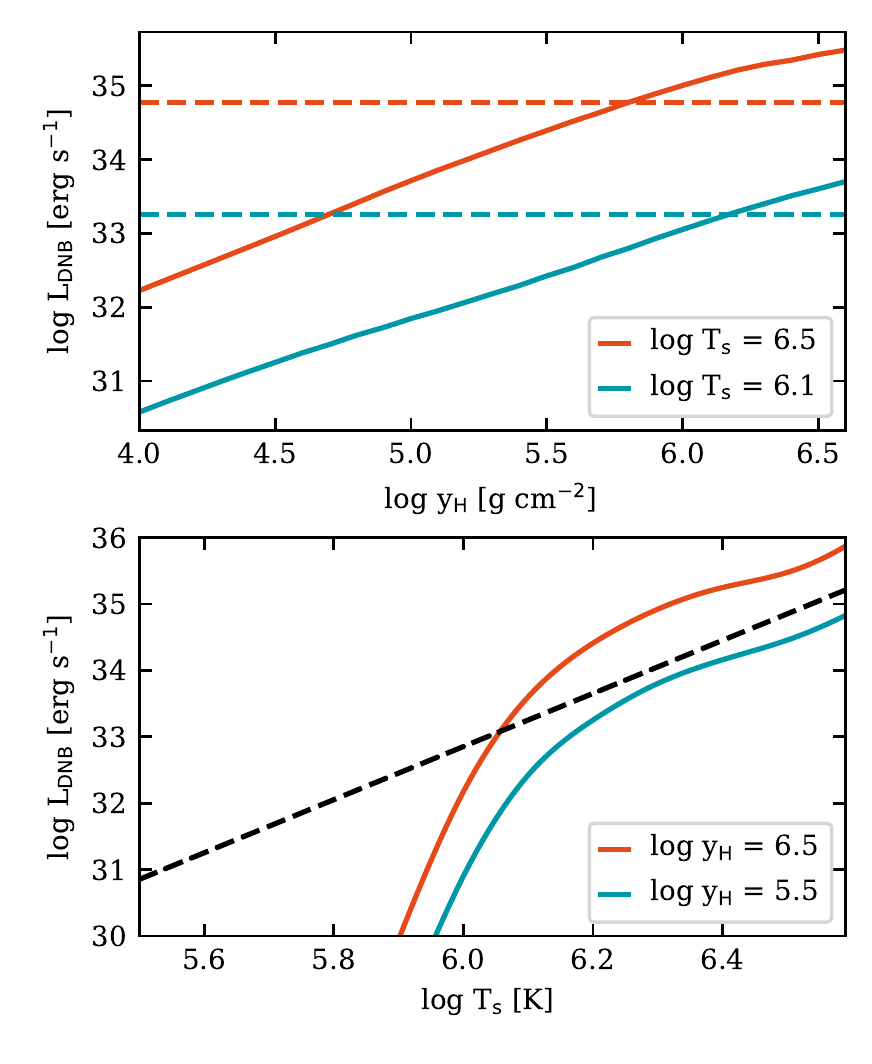}	
	\vspace{-20pt}
	\caption{The DNB luminosities in a H-C envelope, for varying column densities (top) and temperatures (bottom). The luminosity corresponding to the surface temperatures are shown as dashed lines for comparison. }
	\label{fig:HC_dnb_lum}
\end{figure}

In Figure \ref{fig:lums} we show conditions ($\Ts$, $y_{\text{light}}$) when the nuclear burning luminosity becomes relevant by computing the ratio $L_{\text{DNB}}/L_{\text{int}}$ for $y_{\text{light}}$ = \yh~ (left) and $y_{\text{light}}$ = $y_{\text{He}}$ (right). For H-C envelopes, the dominant nuclear reactions are proton captures, which efficiently produce heat for a relatively large part of the \yh~-\Ts~parameter space. For surface temperatures above $\sim$ 10$^6$ K and hydrogen columns larger than $\sim$10$^{5}$ g \pcm, the energy deposited by nuclear burning becomes comparable to and larger than the interior luminosity (see left panel in Figure \ref{fig:lums}). The hydrogen column densities and surface temperatures where the DNB luminosity becomes non-negligible are in the range inferred from observations of neutron stars in LMXBs after accretion outbursts (see \citealt{2017JApA...38...49W} for an observational overview). In Figure \ref{fig:HC_dnb_lum} we show the magnitude of the nuclear burning luminosity for varying hydrogen column (for~\Ts~=~1.26~$\times$~10$^{6}$~K and \Ts = 3.16 $\times$ 10$^{6}$ K) in the top panel, and varying surface temperature in the bottom panel (for \yh = 3.16 $\times$ 10$^{4}$ g \pcm~and \yh = 3.16 $\times$ 10$^{6}$ g \pcm). As an initially large hydrogen column after an accretion episode is not unlikely, the energy deposited in the envelope by diffusive nuclear burning could affect the surface temperature evolution after the outburst.

\begin{figure}
	% To include a figure from a file named example.*
	% Allowable file formats are eps or ps if compiling using latex
	% or pdf, png, jpg if compiling using pdflatex
	%	 Canonical_temp_Teff_tburst_1yr_mdot_1e17.dat
	\includegraphics[width=\linewidth]{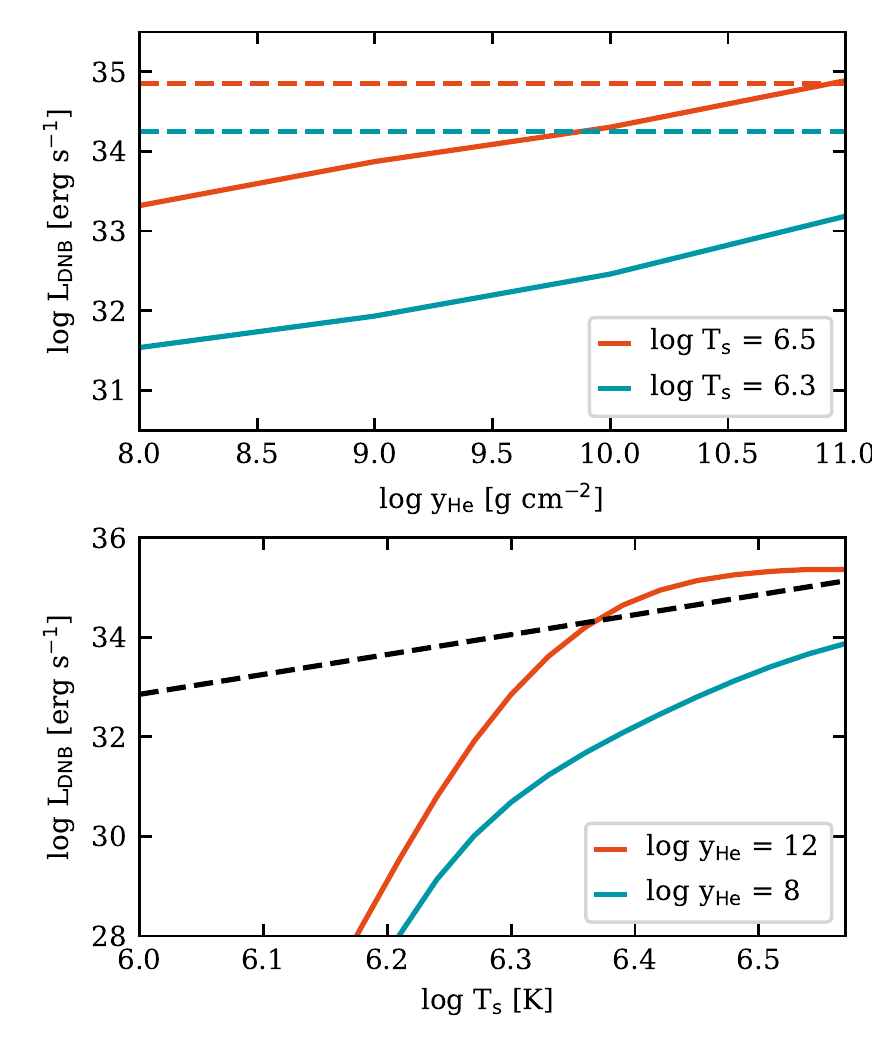}	
	\vspace{-20pt}
	\caption{The DNB luminosities in a He-C envelope, for varying column densities (top) and temperatures (bottom). The luminosity corresponding to the surface temperatures are shown as dashed lines for comparison. }
	\label{fig:HeC_dnb_lum}
\end{figure}

For He-C envelopes, where the dominant nuclear reaction are highly temperature sensitive alpha-captures, the heat generated by nuclear reactions is only relevant for very high temperatures, \Ts > 2 $\times$ 10$^6$ K, and helium columns, $y_{\text{He}} \gtrsim 10^9$ g \pcm, as is shown in the right panel of Figure \ref{fig:lums}. In this regime, the luminosity from nuclear burning becomes larger than $10^{34}$ erg s$^{-1}$, as is shown in Figure \ref{fig:HeC_dnb_lum}. For smaller columns and temperatures, the nuclear burning luminosity rapidly drops and is negligible compared to the interior luminosity (indicated by the dashed lines). The nuclear burning luminosity is relevant compared to the interior luminosity (indicated by the dashed lines) for a smaller range of (and at significantly higher) surface temperatures and column densities compared to H-C envelopes.

%\section{Heating after accretion outbursts (in quiescence)}
\section{DNB Effect on post-outburst cooling curves}
\label{sec:effectcurves}

We investigate the heating effect of DNB on the post-outburst temperature evolution of a cooling NS crust. We simulate the accretion outburst to obtain the post-outburst interior luminosity evolution using the relativistic cooling code \NSCool~\citep{2016ascl.soft09009P} for a NS with mass $M$~=~1.4~$\Msun$ and radius $R$~=~11.5 km. The outer boundary condition (at the bottom of the envelope, $\rhob$) of the cooling code is that the interior luminosity at $\rhob$ is the same as the luminosity at the surface (i.e., no energy is generated or lost in the envelope): $L_\text{s}$ = $L_{\text{int}}$. Thus, we obtain the surface temperature without DNB ($\Tso$) directly from $L_{\text{int}}$. In the following, we will assume the heat generated by DNB in the low density envelope flows towards the surface and DNB does not affect the interior luminosity. Therefore, we can calculate the heating effect of DNB by post-processing the interior luminosity and surface temperature evolution from \NSCool.

We convert the surface temperature without DNB ($\Tso$) to a surface temperature with DNB ($\Ts$) using $L_\text{s}$~=~$L_{\text{int}}~+~x_{\text{out}}L_{\text{DNB}}$, where $L_{\text{DNB}}$ is the DNB luminosity and we introduce $x_{\text{out}}$ as the fraction of DNB luminosity that is radiated outwards so that it is easy to investigate the effect of relaxing our assumption of $x_\text{{out}}$ = 1. The resulting surface temperature including the heat generated by DNB can then be calculated as

\begin{align}
\Ts = \Tso \left(1 + \frac{x_{\text{out}} L_{\text{DNB}}}{L_{\text{int}}} \right)^{1/4}. %= \Tso \left(1 + \alpha \right)^{1/4}.
\end{align}

For a given $M$ and $R$, the nuclear burning luminosity for a H-C envelope can be calculated as a function of $\Tb$ and $\yh$ (analogue to Figure \ref{fig:lums}). Before we can calculate the nuclear burning luminosity for a given \Tb, we convert the surface temperature, $\Tso$, to the corresponding \Tb~using the analytic envelope relations for a H-C envelope described in Section \ref{sec:tstb}.

\subsection{Effect on cooling curves due to heat generation}
\label{sec:effectheating}

\begin{figure}
	% To include a figure from a file named example.*
	% Allowable file formats are eps or ps if compiling using latex
	% or pdf, png, jpg if compiling using pdflatex
	\includegraphics[width=\linewidth]{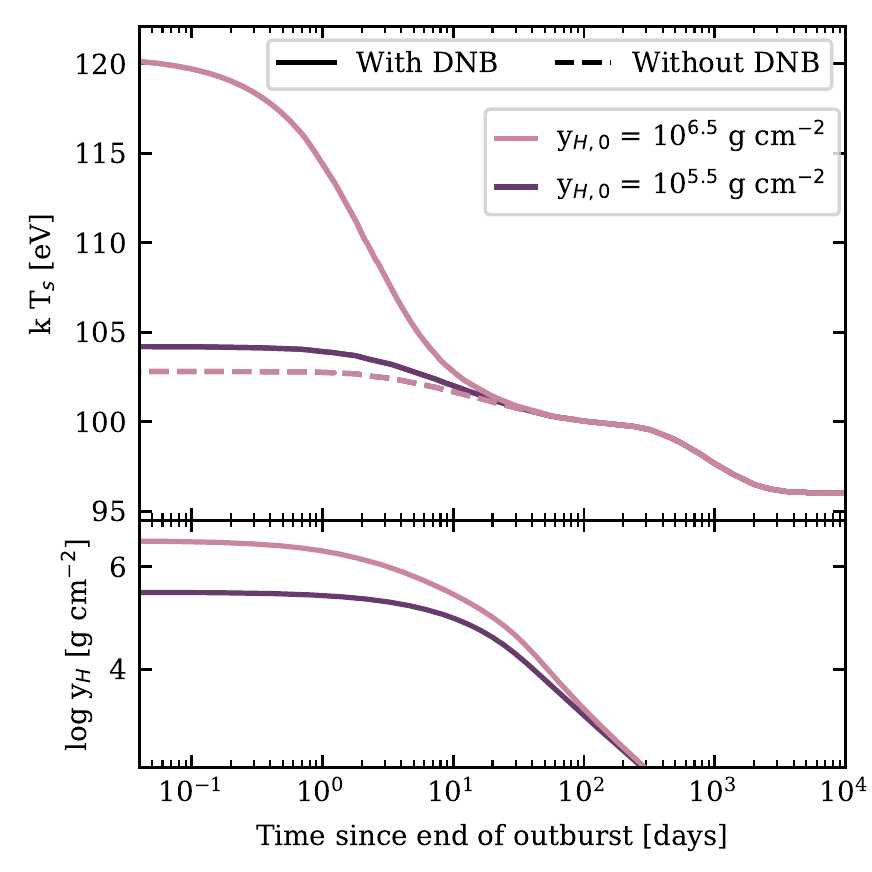}
	\vspace{-20pt}
	\caption{Comparison of the post-outburst surface temperature evolution with (solid) and without (dashed) additional DNB heating for different initial hydrogen columns (see line colours). The interior luminosity was calculated using an outburst duration of 1 year and $\dot{M}$~=~10$^{17}$~g~s$^{-1}$ $\approx$ 2 $\times 10^{-9} \Msun$ yr$^{-1}$.}
	\label{fig:coolingcomp1}
\end{figure}

\begin{figure}
	% To include a figure from a file named example.*
	% Allowable file formats are eps or ps if compiling using latex
	% or pdf, png, jpg if compiling using pdflatex
	\includegraphics[width=\linewidth]{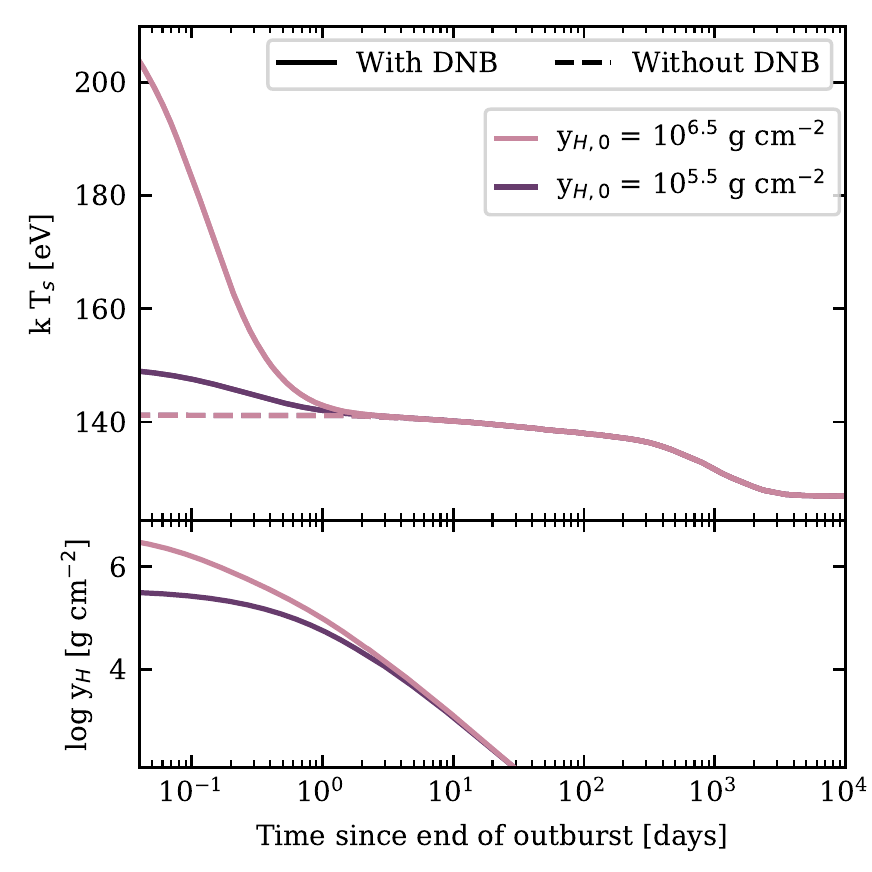}
	\vspace{-20pt}
	\caption{Comparison of the post-outburst surface temperature evolution with (solid) and without (dashed) additional DNB heating for different initial hydrogen columns (see line colours). The interior luminosity was calculated using an outburst duration of 10 years and $\dot{M}$~=~10$^{17}$~g~s$^{-1}$ $\approx$ 2 $\times 10^{-9} \Msun$ yr$^{-1}$.}
	\label{fig:coolingcomp2}
\end{figure}

In Figure \ref{fig:coolingcomp1} and \ref{fig:coolingcomp2}, we show the resulting surface temperature evolution for an accretion outburst duration of 1 and 10 years, respectively, with constant accretion rates. As the amount of heat deposited on the star during the outburst is sensitive to the outburst duration and accretion rate, we vary either one of these parameters to obtain varying interior luminosity strengths after the outburst. We compare the surface temperature with and without DNB, for two initial hydrogen columns and use the burning rate at each timestep to evolve the hydrogen column size.

Figures \ref{fig:coolingcomp1} and \ref{fig:coolingcomp2} clearly show that the importance of DNB depends sensitively on the initial hydrogen column after the accretion outburst, as this sets the amount of hydrogen available for burning. For initial hydrogen columns $< 3.2 \times 10^{5}$ g \pcm, the change in surface temperature due to DNB is $<$~5\%. For larger initial hydrogen columns (\yh~$\gtrsim$ 3.2 $\times$ 10$^{6}$ g \pcm), which are typically assumed in cooling studies for a solar mixture of light elements, the initial change in surface temperature is $>$5 - 50 \% depending on the interior luminosity strength set by the accretion history. In the following, we discuss the low and high post-outburst luminosity scenarios for the case when the cooling starts with a large initial hydrogen column (log \yh [g cm$^{-2}$] = 6.5).

Figure \ref{fig:coolingcomp1} corresponds to the scenario of a relatively short accretion period, resulting in low post-outburst cooling luminosities and surface temperatures. Despite the relatively low temperatures, $T_\text{{s,old}}^{0}$ = 1.2 $\times 10^6$ K (corresponding to $k \Ts$ = 103 eV), when there is a substantial hydrogen column, the heating due to DNB is large and increases the initial post-outburst surface temperature by $\sim$19$\%$ to $\Ts^0 \sim$ 1.4 $\times$ 10$^6$ K (corresponding to $k \Ts$ = 120 eV). The absolute increase in initial surface temperature is $\Delta T_{\text{DNB}}$ = 2.2 $\times$~10$^5$~K. For these temperatures, the duration of higher surface temperatures due to heat released by DNB is $\sim$ 10 days.

For longer accretion outbursts, which result in higher post-outburst luminosities (see Figure \ref{fig:coolingcomp2}), the initial increase in surface temperature due to DNB is larger (when the initial hydrogen column is large enough, $\yh \gtrsim 1.2 \times 10^{6}$ g \pcm). In this case, the initial surface temperature increases by 47\% corresponding to an absolute increase in temperature of $\Delta T_{\text{DNB}}$ = 7.7 $\times$ 10$^5$ K. This means that the initial surface temperature has increased from $T_\text{{s,old}}^{0}$ = 1.6 $\times 10^6$ K (corresponding to $k \Ts$ = 141 eV), to $Ts^0 \sim$ 2.4 $\times$ 10$^6$ K (corresponding to $k \Ts$ = 204 eV). However, higher post-outburst luminosities lead to shorter durations ($\sim$hours to days) when DNB contributes significantly to the total luminosity and surface temperature. The total amount of energy available from burning hydrogen is set by the initial hydrogen column, and corresponds to the integral of the elevated luminosity cooling curves, which is constant in all cases with the same initial hydrogen column. For the initial hydrogen column of  $\yh \sim 1.2 \times 10^{6}$ g \pcm, the total amount of energy that is released from hydrogen burning is 10$^{38}$ erg (i.e., $\sim$7 MeV per nucleon).

Without residual accretion or a large helium buffer that may slow down the depletion of hydrogen, the times when DNB significantly increases the observed surface temperature for a H-C envelope are $\sim$ 0.1 - 80 days from the end of the accretion outburst, depending on the initial hydrogen column and cooling luminosity. For larger initial cooling luminosities (and thus higher surface temperatures), the initial change in surface temperatures increases, while the relevant duration decreases, as the available hydrogen is consumed more rapidly. 

\subsection{Effect on cooling curves due to changing composition}
\label{sec:effectcomposition}

\begin{figure}
	% To include a figure from a file named example.*
	% Allowable file formats are eps or ps if compiling using latex
	% or pdf, png, jpg if compiling using pdflatex
%
	\includegraphics[width=\linewidth]{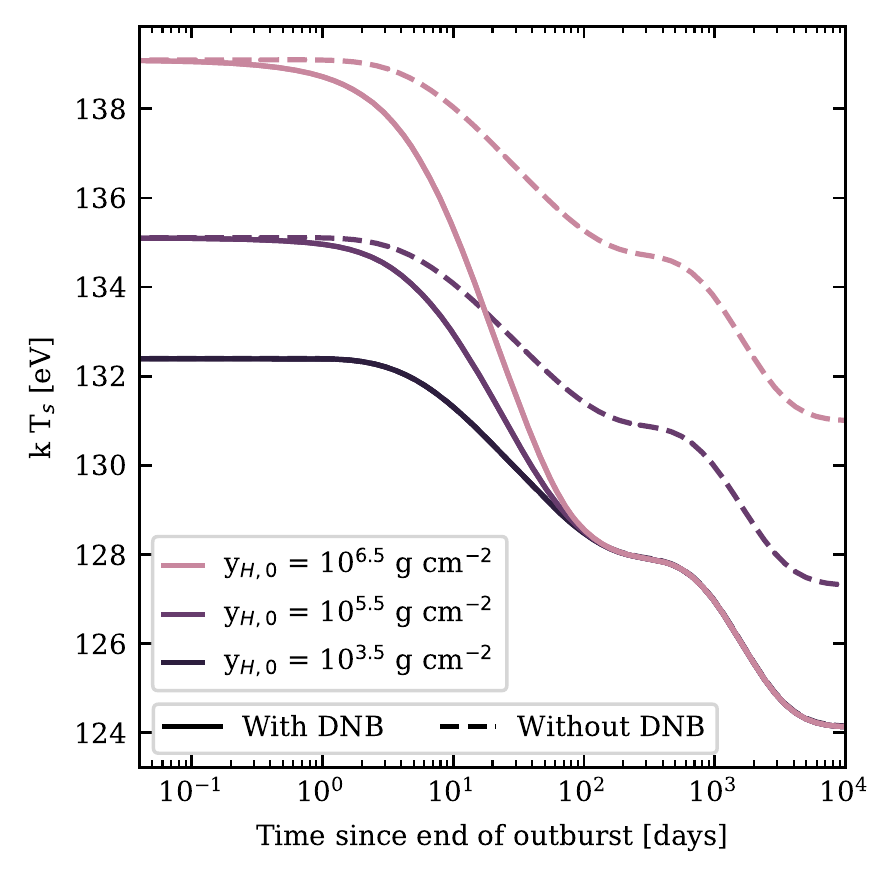}
	\vspace{-20pt}
	\caption{The effect of a changing hydrogen column due to DNB (while ignoring heat release, see text) on the the post-outburst surface temperature evolution for different (initial) hydrogen columns (see linecolors). The interior cooling was calculated using an outburst duration of 1 years and $\dot{M}$~=~10$^{17}$~g~s$^{-1}$ $\approx$ 2 $\times 10^{-9} \Msun$ yr$^{-1}$.}
	\label{fig:cooling-effect-tstb}
\end{figure}

In addition to releasing energy through nuclear reactions, DNB can alter the observed cooling after accretion outbursts by altering the light element column in the envelope over time. In Section \ref{sec:tstb}, we show the effect on the \Ts-\Tb~temperature relations for varying hydrogen column sizes in H-C envelopes (see Figure \ref{fig:HC_TsTb}). The decreasing hydrogen column due to DNB, will only affect the cooling curve when the hydrogen column is varied in the transition region (as shown in Figure \ref{fig:HCtransition}). At the relevant temperatures for NSs in quiescence (with initial post-outburst surface temperatures~\Ts~$>$ 8 $\times$ 10$^5$ K), this means that the same boundary temperature (at the bottom of the envelope) corresponds to different surface temperatures when the hydrogen column varies between $10^{4}$ g \pcm $\lesssim \yh \lesssim 10^{7}$ g \pcm. When the initial hydrogen column after the accretion outburst is smaller than \yh $\sim 10^{4}$ g \pcm, a further decreasing hydrogen column will not affect the cooling curve. When the initial hydrogen column is larger than \yh $\sim 10^{4}$ g \pcm, the cooling curve transitions from relatively hydrogen-rich envelope temperature relations towards those for a hydrogen-poor envelope. We show this effect in Figure \ref{fig:cooling-effect-tstb}, where we ignore the effect of heat released by DNB and only show the effect of a changing envelope composition. The dashed cooling curves in Figure \ref{fig:cooling-effect-tstb} correspond to fixed hydrogen columns, while the solid cooling curves correspond to the scenario where the hydrogen column decreases over time due to DNB. When DNB is taken into account, all the plotted cooling curves (irrespective of initial column density) correspond to that of a hydrogen-poor envelope after $\sim$ 50 days. Note that for larger post-outburst luminosities, it can take less time ($\sim$ 1 day) to decrease the hydrogen column below \yh $\sim 10^{4}$ g \pcm ~(see, e.g., the bottom panel in Figure \ref{fig:coolingcomp2}).  

Currently, the unknown post-outburst envelope composition introduces an uncertainty in cooling models through the \Ts-\Tb~relations and is often left as a free fit variable. By taking into account DNB, it is possible to further constrain the post-outburst envelope composition, although the initial composition that is left after the accretion outburst is still uncertain. Regardless of the initial hydrogen column, we find that our cooling curves correspond to the hydrogen-poor limit within $\sim$100 days. 

\section{Application to crust cooling sources}
\label{sec:sources}

In this section, we calculate cooling curves including DNB for observed crust-cooling sources in LMXBs. We select sources for which shallow heating was invoked in order to explain the observed temperatures (see Section \ref{sec:introduction}). We model the sources without shallow heating to show how the quiescent heating due to DNB compares to observations. We use the method described in Section \ref{sec:effectcurves}, but include the time-variable accretion rate described in \cite{2016MNRAS.461.4400O}  to estimate the effect of DNB luminosity on cooling curves.

\subsection{Aql X-1}

\begin{figure}
	% To include a figure from a file named example.*
	% Allowable file formats are eps or ps if compiling using latex
	% or pdf, png, jpg if compiling using pdflatex
	%	\includegraphics[width=\linewidth]{figures/Terzan5X-2_DNB_CoolingCurves.pdf}
	\includegraphics[width=\linewidth]{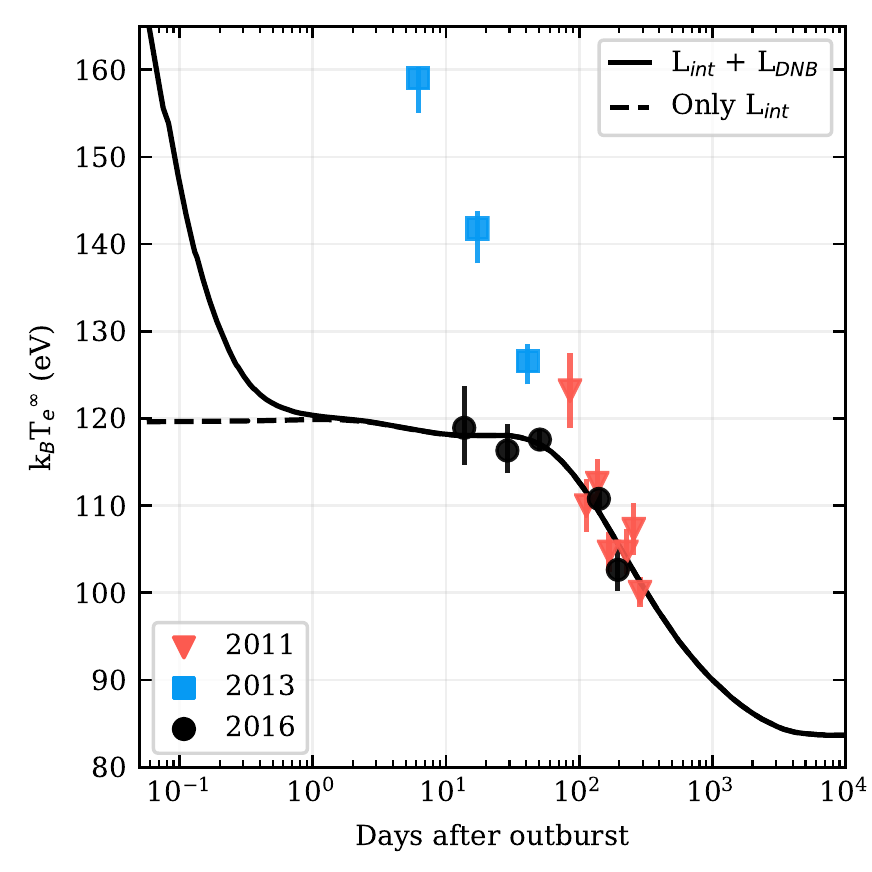}
	\vspace{-20pt}
	\caption{Illustrative cooling curves for the neutron star in Aql X-1 compared to the inferred temperatures after its 2011 (red triangles), 2013 (blue squares) and 2016 (black circles) outbursts. The cooling curves in this graph correspond to the same cooling model with the same NS properties. The differences between the curves are due the absence/presence of DNB. The dashed line shows a cooling model that includes a shallow heat source with Q$_{\text{shallow}}$ = 1.3 MeV nuc$^{-1}$. The solid line shows the same cooling model when heating due to DNB is included. As the hydrogen column is quickly depleted at these temperatures, the DNB luminosity is only relevant $<$ 1 day. }
	\label{fig:AqlX1}
\end{figure}

The NS in the Aql X-1 LMXB displays frequent accretion outbursts with typical durations of a few months (see, e.g., \citealt{2016MNRAS.456.4001W}). Due to its frequent outbursts, Aql X-1 is a promising source for the study of shallow heating during outbursts. \cite{2018MNRAS.477.2900O} modelled its thermal evolution for the period between 1996-2015, which includes 23 accretion outbursts. They found that the observed cooling after multiple outbursts could be explained using different magnitudes and depths of the shallow heating source. New temperature data for the cooling after its 2016 outburst was presented by \cite{2019arXiv190707205D}, who further investigated the shallow heating properties by comparing the cooling after its 2011, 2013 and 2016 outbursts. The three outbursts showed a striking similarity in outburst properties such as duration, peak flux and overall shape, but were followed by cooling tracks with very distinct differences in the early time thermal evolution. \cite{2019arXiv190707205D} found that the depth and magnitude of the shallow heating source during the 2016 outburst must have been larger than during the 2013 outburst\footnote{No strong constraints could be placed on the shallow heating properties during the 2011 outburst due to the lack of observations during the early cooling phase $<$80 days \citep{2018MNRAS.477.2900O,2019arXiv190707205D}.}. This indicates that the shallow heating properties can be different after multiple similar outbursts in the same source. Here, we briefly explore an alternative scenario for the differences in the early time cooling curves by calculating the heating effect from DNB of accreted hydrogen.

We use \NSCool~to calculate an illustrative cooling model that loosely follows the inferred temperatures after the 2016 outburst as well as the late-time cooling of the 2011 and 2013 outbursts. We do not attempt to find a best-fit model, as we only intend to investigate whether a cooling model with the same NS, accretion- and shallow heating properties could explain the observed differences at early times. We use a mass and radius of M = 1.6 $\Msun$ and R = 11 km, which are consistent with those used in the spectral fits to obtain the temperature data \citep{2018MNRAS.477.2900O,2019arXiv190707205D}. As the three outbursts had similar durations and accretion rates, we use an average accretion rate of $\dot{M}$~=~10$^{18}$~g~s$^{-1}$ for an outburst duration of 2.5 months. Our model assumes a core temperature of T$_{\text{core}}$~=~1.1~$\times$~10$^{8}$~K and includes a shallow heating source of Q$_{\text{shallow}}$ = 1.3 MeV nuc$^{-1}$ at a maximum depth of $\rho$ = 2.5 $\times$ 10$^{10}$ g cm$^{-3}$.

In Figure \ref{fig:AqlX1}, we compare the inferred temperatures after the three outbursts to the cooling track given by our model (dashed line) as well as the track given by the same model when luminosity from DNB is added (solid line). When including the DNB luminosity, we assume an initial hydrogen column after the outburst of \yh = 1.2 $\times$ 10$^{6}$ g \pcm. We find that, due to the high temperatures, the hydrogen column drops rapidly to  \yh = 3.2 $\times$ 10$^{4}$ g \pcm~within 1~day and no significant luminosity is produced by DNB after that time. Therefore, the cooling curves with and without a DNB luminosity are the same after 1~day and DNB of hydrogen alone does not explain the observed differences up to 80 days into the cooling. It is also unlikely that the differences in early time cooling between the outbursts can be explained with DNB and residual low-level accretion, as the accretion rates required produce a luminosity that would dominate the total observed luminosity.

Note that in all our cooling curves including DNB, we made the assumption that the generated heat immediately leaves the star at the surface and has no effect on the interior cooling. Therefore, the cooling curves are only affected during active DNB. An inwards heatflow could potentially prolong the timescale over which DNB affects the cooling curve when accretion is not included, as the heat will reach the surface at later times. However, as the heat is produced at shallow depths, this effect is expected to be small.

\subsection{MAXI J0556-332}
\label{sec:maxi}

The cooling after the 2012 outburst of MAXI J0556-332 is the most extreme case for which shallow heating has been invoked. The temperatures after the end of the outburst, are by far the hottest ever observed. A very large amount of shallow heating, $\sim$15 MeV nuc$^{-1}$, was necessary during the outburst to explain the observed cooling after the outburst was over \citep{2014ApJ...795..131H,2015ApJ...809L..31D,2017ApJ...851L..28P}, which is about an order of magnitude more than what was needed for other sources. Interestingly, for the cooling after two subsequent reheating outbursts in this source, much less shallow heating was needed \citep{2017ApJ...851L..28P}.

We discuss DNB in relation to this source, as it poses the most extreme scenario in terms of both shallow heating and large post-outburst temperatures. In fact, the source is so hot that besides proton captures onto carbon, another nuclear reaction needs to be included. At temperatures below 10$^{8}$ K, a result of the proton capture onto C is the creation of $^{13}$N, which decays into $^{13}$C. At temperatures $T > 10^{8}$ K, the time in which $^{13}$N captures hydrogen is shorter than the time in which it decays, thus adding another reaction in which additional heat is released. Therefore, we also include the second step of the hot CNO cycle, such that the hydrogen capturing reactions are:

\begin{align}
{\rm H}^{1}_{1} + {\rm C}^{12}_{6} \rightarrow {\rm N}^{13}_{7} + \gamma, \\
{\rm H}^{1}_{1} + {\rm N}^{13}_{7} \rightarrow {\rm O}^{14}_{8} + \gamma,
\end{align}

\noindent
where the second reaction deposits an additional amount of energy in the envelope (i.e, 4.63 MeV per reaction). \\

We find that, even with the inclusion of hydrogen captures onto nitrogen, DNB alone does not produce enough heat by far to explain the large observed temperatures observed in MAXI J0556-332. At these temperatures, the burning rate is so high that large accretion rates are required to maintain the hydrogen column at levels at which enough heat is produced. Even with the inclusion of decaying low-level accretion rates after the outburst, we find that very large initial accretion rates ($\dot{M}_{\text{init}} \sim 10^{15}$ g s$^{-1}$) are required to reach the observed post-outburst temperatures. These accretion rates can be ruled out, as the corresponding accretion luminosity ($\sim 10^{35}$ erg s$^{-1}$) would be significantly larger than the observed luminosity. We also consider a possible He-C envelope, as the conditions are suitable for helium burning. However, we find that while a large helium column (y$_{\rm He} \sim $ 10$^{12}$ g \pcm) will increase the temperature during the full cooling time, the shift is not large enough to explain the observed temperatures. Smaller helium columns will not make a significant contribution to heat generation (see luminosity in Figure \ref{fig:lums}).

\begin{figure}
	% To include a figure from a file named example.*
	% Allowable file formats are eps or ps if compiling using latex
	% or pdf, png, jpg if compiling using pdflatex
	\includegraphics[width=\linewidth]{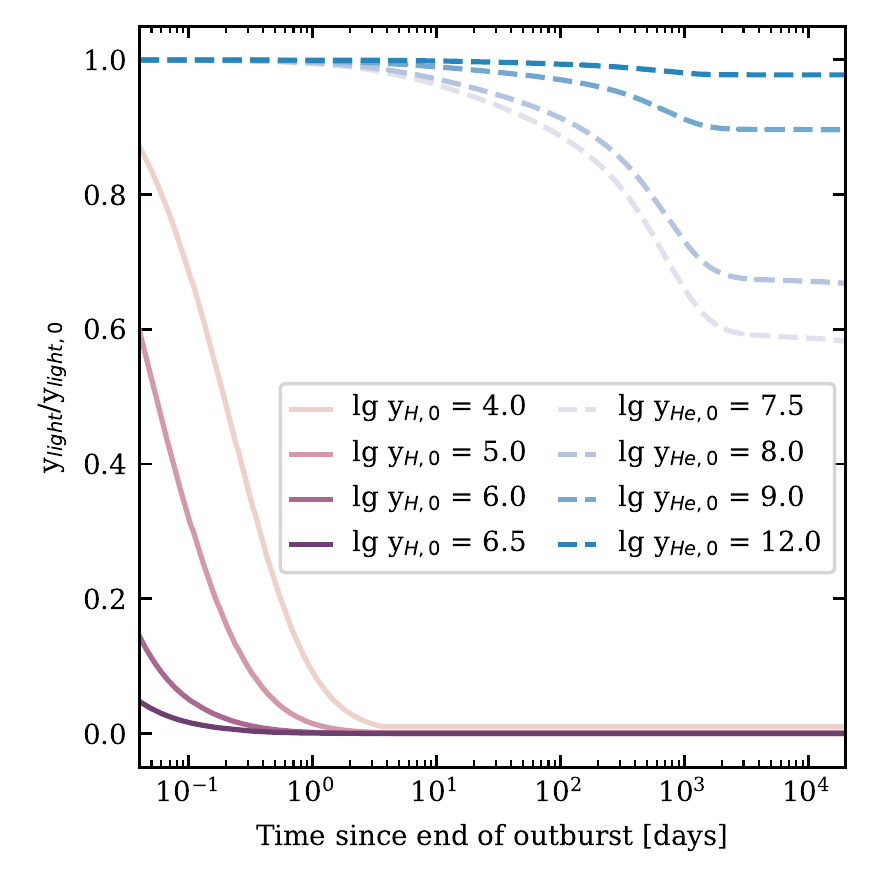}	
	\vspace{-20pt}
	\caption{Evolution of the light element column through DNB of MAXI J0556-332 for hydrogen (in a H-C envelope) and helium (in a He-C envelope) for varying initial column sizes. All initial hydrogen is consumed in the H-C envelope, while initially large helium column sizes decrease but remain present despite the relatively large temperatures. }
	\label{fig:MAXI_composition}
\end{figure}

In Figure \ref{fig:MAXI_composition}, we show the relative change in column size for varying initial light element columns. For H-C envelopes, we find that all hydrogen is rapidly depleted within $\sim$20 days, regardless of the initial column size after the accretion outburst. For He-C envelopes, where only helium that penetrates deep in the envelope is consumed by captures onto heavier elements, the helium column can decrease but is not completely consumed in the crust-cooling time. It is interesting that observations of a hydrogen atmosphere in this source, may be evidence for the presence of low-level accretion in quiescence as all initial hydrogen would be consumed within 20 days. Note that only a small accretion rate is needed to regain a hydrogen atmosphere (see \citealt{2004ApJ...605..830C,2019MNRAS.484..974W} for a discussion on the effect of DNB on the composition of the atmosphere). 

We note that other nuclear reactions may be taking place which are currently unaccounted for. \cite{2017JPSJ...86l3901L} calculate cooling curves for MAXI J0556-332 without shallow heating but including the full beta-decay limited hot CNO cycle and while assuming a decaying accretion rate during quiescence. In their calculation, nuclear burning does not occur diffusively but follows a fixed burning rate for the hot CNO cycle (independent of the amount of hydrogen available and the density) in the region of the envelope where $T > 10^9$ K and changes in abundances are not taken into account. They conclude that the hot CNO cycle can significantly increase the surface temperature and find that the hot CNO cycle operates up to $\sim$500 days into quiescence. However, as we show here, when changes in abundance are taken into account (i.e., hydrogen depletion due to nuclear burning) the increase in luminosity due to hydrogen burning effectively shuts off much sooner (within $20$ days if no further accretion takes place). 

\section{Discussion}

\subsection{General effects of DNB}
We study the effect of diffusive nuclear burning on observed neutron star crust cooling in three ways: 1) by obtaining static temperature relations for H-C envelopes that include DNB (Section \ref{sec:tstb}), 2) by considering the additional heating due to DNB (Section \ref{sec:effectheating}), and 3) by considering time variable light element columns (Section \ref{sec:effectcomposition}). It is typically assumed that no heat is generated in the NS envelope during quiescence, but here we find that the DNB luminosity is not always negligible. Specifically, the DNB luminosity is relevant in H-C envelopes for hydrogen columns $\gtrsim 10^{5}$ g \pcm~at surface temperatures $\gtrsim$ 1 MK, and for He-C envelopes for helium columns $\gtrsim 10^{11}$ g \pcm~at surface temperatures $\gtrsim$ 2 MK. Thus, the surface luminosity can be dominated by DNB for relevant parts of the parameter space (see Figure \ref{fig:lums}) which can affect the interpretation of early time surface temperature observations. However, as the light element column decreases rapidly, the DNB luminosity is only relevant at times within $\sim$80 days after the outburst for low post-outburst temperatures and within $\sim$1 day for larger post-outburst temperatures (see Section \ref{fig:HC_dnb_lum}).

We find that the conditions in quiescent NSs in LMXBs are such that active DNB can rapidly decrease the light element column size on relatively short timescales (within $\sim$1-100 days). Good data sampling in the early phases (<50-100 days) are necessary to see the effects. This result can be relevant for crust cooling studies, as the \Ts-\Tb~relations are not fixed for one column size during the time the crust cools back to thermal equilibrium with the core. Instead, the \Ts-\Tb~relations change accordingly when the light element column size decreases in the transition region (see Figure \ref{fig:HCtransition}). This could manifest in two ways: 

\begin{itemize}
	\item For cooler post-outburst temperatures, the hydrogen column is consumed at a lower rate and the corresponding change in \Ts-\Tb~relations is potentially observable as an initial decrease in surface temperature at $\lesssim$ 80 days after the end of the outburst (see Figure \ref{fig:cooling-effect-tstb}). This initial cooling is dominated by the change in composition and thus transparency of the envelope, rather than cooling of the interior. 
	\item For hotter post-outburst temperatures, the hydrogen column is consumed at a sufficiently short timescale (within hours to a few days) such that the full observable cooling curve corresponds to a static, hydrogen-poor envelope. Limits on the envelope composition due to rapid DNB at early times can be used to constrain the static envelope composition at later times if the post-outburst accretion rate is assumed to be negligible for replenishing the light element column in the sensitivity strip.
\end{itemize}
 
Our main conclusion is that the current static treatment of the neutron star envelope during thermal evolution studies, even on the short timescales considered in cooling studies, may not be appropriate. Processes such as DNB can significantly alter the envelope composition over time and can release additional heat, making it worthwhile to include the burning region of the envelope in future thermal evolution calculations. This would require setting the upper boundary of thermal transport equations to lower densities than $\rhob$= 10$^{8}$ or $\rhob$= 10$^{10}$ instead of using static temperature relations to connect the temperature at those densities to the temperature at the surface (see \citealt{2020ApJ...888...97B} for a more consistent treatment of the envelope applied to young neutron stars). 

\subsection{DNB as shallow heating mechanism?}

We find that DNB can increase the initial post outburst surface temperatures to even larger values than the observed initial temperature, without invoking an unknown shallow heat source. However, the temperature increase due to DNB drops too rapidly ($<$10 days) to explain the large observed surface temperatures at 50-100 days for Aql X-1. We find that without an additional shallow heating source, large accretion rates are necessary to replenish the hydrogen column which would produce an X-ray luminosity that would dominate the total luminosity. Therefore, DNB of hydrogen alone is not able to lift the need for an additional shallow heating source during the outbursts of Aql X-1. DNB in both H-C and He-C envelopes cannot replace shallow heating for MAXI J0556-332, the source with the most extreme post-outburst temperatures (see Section \ref{sec:maxi}). The large post-outburst accretion rates needed to match these temperatures with DNB, would overwhelm diffusive burning and produce accretion luminosities that are larger than the observed luminosity. 

The amount of shallow heating has been found to differ between sources and even after multiple outbursts in the same source. Nuclear burning could generate different amounts of additional heating for different post-outburst envelope compositions (which depend on the accretion history). We examine whether including both DNB and shallow heating for multiple Aql X-1 outbursts (i.e., the 2011, 2013 and 2016 outbursts from \citealt{2018MNRAS.477.2900O} and \citealt{2019arXiv190707205D}) could lift the need for a varying amount of shallow heating between outbursts. We find that at the inferred post-outburst temperatures, the hydrogen in the envelope would be consumed within 1 day and DNB no longer produces a significant luminosity at the time of the cooling observations. Furthermore, the required accretion rates to replenish the hydrogen column sufficiently to explain the early time cooling observations are expected to produce accretion luminosities much larger than the observed luminosity.

It would be interesting to include nuclear burning in the envelope both during the accretion outburst (non-diffusive burning) as well as in quiescence (diffusive burning), as potential heat flow into the neutron star interior is currently not accounted for and could extend the time when DNB is relevant. The work presented here involves calculating independently the evolutions of interior luminosity, DNB luminosity and size of light element column. It would be worthwhile to perform a self-consistent calculation, which includes a cooling code outer boundary condition that can vary with the burning region. Such a study could also include a full nuclear reaction network to evolve the envelope composition beyond two component models considered here and in previous works.
 
\section*{Acknowledgements}

MJPW thanks the organisers and participants of the ICONS 2019 meeting in Amsterdam for useful presentations and discussions that contributed to this paper. WCGH acknowledges support through grant ST$/$R00045X$/$1 from Science and Technology Facilities Council in the UK. We acknowledge the support of the PHAROS COST Action (CA16214). LSO and RW are supported by an NWO Top Grant, module 1, awarded to RW. DP and MB acknowledge financial support by the Mexican Consejo Nacional de Ciencia y Tecnolog$í$a with a CB-2014-1 grant \#240512. MB also acknowledges support from a postdoctoral fellowship from UNAM-DGAPA.

\appendix

\section{Analytic T$_\text{s}$-T$_\text{b}$ relations}
\label{sec:appendix_tbts}

\begin{center}
	\begin{table}
		\caption{Best fit parameters for the H-C mixture with $\rhob$ = 10$^{10}$ g cm$^{-3}$  (Equations~\ref{eq:fitsubeq}).}
		\label{tab:fitparams-HeC}
		\begin{tabular}{c c c c c c} % four columns, alignment for each
			\hline
			$p_1$ & $p_2$ & $p_3$ & $p_4$ & $p_5$ & $p_6$ \\
			\hline
			5.708 & 0.09519 & 1.667 & 4.676 & 0.05793 & 1.705 \\
			\\
			$p_7$ & $p_8$ & $p_9$ & $p_{10}$ & $p_{11}$ & \\
			\hline
			16710 & 3.178 & 4.681 & 1.719 & 0.9215 &  \\
			\hline
		\end{tabular}	
	\end{table}
\end{center}

In this section, we present accurate fits to the computed $\Ts$-$\Tb$-$\yh$ data which can be used in cooling simulations. All the fits are obtained for an envelope with surface gravity $g_\text{{s,0}}$ = 2.4271 $\times$ 10$^{14}$ cm s$^{-2}$ but can be scaled for any $g_\text{s}$ using $Y = (\Ts/1{\rm MK})(g_\text{{s,0}}/g_\text{s})^{1/4}$ \citep{1983ApJ...272..286G}. We use an adapted version of the analytic functions (Equation \ref{eq:fiteq}) presented by \cite{2016MNRAS.459.1569B}.

\begin{equation}
\begin{split}
\label{eq:fiteq}
T_b (Y, \yh) = 10^7 K \times \Biggl( f_4(Y) + [f_1(Y) - f_4(Y)] \\
\times \left[1 + \left( \frac{\yh}{f_2(Y)}\right)^{f_3(Y)}\right]^{-0.2} \Biggr)
\end{split}
\end{equation}

The analytic relation fitted for the H-C mixture for an envelope with a lower boundary of $\rhob$ = 10$^{10}$ g cm$^{-3}$ is given by equation \ref{eq:fiteq} which consists of the functions \ref{eq:fitsubeq}. Here, the functions $f_1-f_4$ are identical to those for the He-C mixture in \cite{2016MNRAS.459.1569B}. We obtain a maximum relative error of 0.016. The root mean square of the relative error is 0.0037.

\begin{equation}
\begin{split}
\label{eq:fitsubeq}
f_1(Y) &= p_1Y^{p_2 \text{log}_{10}Y+p_3} \\
f_2(Y) &= p_7 Y^{p_8 (\text{log}_{10} Y)^2 + p_9} \\
f_3(Y) &= p_{10} \sqrt{\frac{Y}{Y^2 + p_{11}^2}} \\
f_4(Y) &= p_4 Y^{p_5 \text{log}_{10} Y + p_6} \\
\end{split}
\end{equation}

%%%%%%%%%%%%%%%%%%%%%%%%%%%%%%%%%%%%%%%%%%%%%%%%%%

%%%%%%%%%%%%%%%%%%%% REFERENCES %%%%%%%%%%%%%%%%%%

% The best way to enter references is to use BibTeX:

\bibliographystyle{mnras}
\bibliography{accretion_references} % if your bibtex file is called example.bib

\begin{thebibliography}{}
\makeatletter
\relax
\def\mn@urlcharsother{\let\do\@makeother \do\$\do\&\do\#\do\^\do\_\do\%\do\~}
\def\mn@doi{\begingroup\mn@urlcharsother \@ifnextchar [ {\mn@doi@}
  {\mn@doi@[]}}
\def\mn@doi@[#1]#2{\def\@tempa{#1}\ifx\@tempa\@empty \href
  {http://dx.doi.org/#2} {doi:#2}\else \href {http://dx.doi.org/#2} {#1}\fi
  \endgroup}
\def\mn@eprint#1#2{\mn@eprint@#1:#2::\@nil}
\def\mn@eprint@arXiv#1{\href {http://arxiv.org/abs/#1} {{\tt arXiv:#1}}}
\def\mn@eprint@dblp#1{\href {http://dblp.uni-trier.de/rec/bibtex/#1.xml}
  {dblp:#1}}
\def\mn@eprint@#1:#2:#3:#4\@nil{\def\@tempa {#1}\def\@tempb {#2}\def\@tempc
  {#3}\ifx \@tempc \@empty \let \@tempc \@tempb \let \@tempb \@tempa \fi \ifx
  \@tempb \@empty \def\@tempb {arXiv}\fi \@ifundefined
  {mn@eprint@\@tempb}{\@tempb:\@tempc}{\expandafter \expandafter \csname
  mn@eprint@\@tempb\endcsname \expandafter{\@tempc}}}

\bibitem[\protect\citeauthoryear{{Beznogov}, {Potekhin}  \&
  {Yakovlev}}{{Beznogov} et~al.}{2016}]{2016MNRAS.459.1569B}
{Beznogov} M.~V.,  {Potekhin} A.~Y.,   {Yakovlev} D.~G.,  2016, \mn@doi
  [\mnras] {10.1093/mnras/stw751}, \href
  {http://adsabs.harvard.edu/abs/2016MNRAS.459.1569B} {459, 1569}

\bibitem[\protect\citeauthoryear{{Beznogov}, {Page}  \&
  {Ramirez-Ruiz}}{{Beznogov} et~al.}{2020}]{2020ApJ...888...97B}
{Beznogov} M.~V.,  {Page} D.,   {Ramirez-Ruiz} E.,  2020, \mn@doi [\apj]
  {10.3847/1538-4357/ab5fd6}, \href
  {https://ui.adsabs.harvard.edu/abs/2020ApJ...888...97B} {888, 97}

\bibitem[\protect\citeauthoryear{{Brown} \& {Cumming}}{{Brown} \&
  {Cumming}}{2009}]{2009ApJ...698.1020B}
{Brown} E.~F.,  {Cumming} A.,  2009, \mn@doi [\apj]
  {10.1088/0004-637X/698/2/1020}, \href
  {https://ui.adsabs.harvard.edu/\#abs/2009ApJ...698.1020B} {698, 1020}

\bibitem[\protect\citeauthoryear{{Brown}, {Bildsten}  \& {Rutledge}}{{Brown}
  et~al.}{1998}]{1998ApJ...504L..95B}
{Brown} E.~F.,  {Bildsten} L.,   {Rutledge} R.~E.,  1998, \mn@doi [\apj]
  {10.1086/311578}, \href
  {https://ui.adsabs.harvard.edu/\#abs/1998ApJ...504L..95B} {504, L95}

\bibitem[\protect\citeauthoryear{{Brown}, {Bildsten}  \& {Chang}}{{Brown}
  et~al.}{2002}]{2002ApJ...574..920B}
{Brown} E.~F.,  {Bildsten} L.,   {Chang} P.,  2002, \mn@doi [\apj]
  {10.1086/341066}, \href {http://adsabs.harvard.edu/abs/2002ApJ...574..920B}
  {574, 920}

\bibitem[\protect\citeauthoryear{{Brown}, {Cumming}, {Fattoyev}, {Horowitz},
  {Page}  \& {Reddy}}{{Brown} et~al.}{2018}]{2018PhRvL.120r2701B}
{Brown} E.~F.,  {Cumming} A.,  {Fattoyev} F.~J.,  {Horowitz} C.~J.,  {Page} D.,
    {Reddy} S.,  2018, \mn@doi [\prl] {10.1103/PhysRevLett.120.182701}, \href
  {https://ui.adsabs.harvard.edu/\#abs/2018PhRvL.120r2701B} {120, 182701}

\bibitem[\protect\citeauthoryear{{Chamel} \& {Haensel}}{{Chamel} \&
  {Haensel}}{2008}]{2008LRR....11...10C}
{Chamel} N.,  {Haensel} P.,  2008, \mn@doi [Living Reviews in Relativity]
  {10.12942/lrr-2008-10}, \href
  {https://ui.adsabs.harvard.edu/abs/2008LRR....11...10C} {11, 10}

\bibitem[\protect\citeauthoryear{{Chang} \& {Bildsten}}{{Chang} \&
  {Bildsten}}{2003}]{2003ApJ...585..464C}
{Chang} P.,  {Bildsten} L.,  2003, \mn@doi [\apj] {10.1086/345551}, \href
  {http://adsabs.harvard.edu/abs/2003ApJ...585..464C} {585, 464}

\bibitem[\protect\citeauthoryear{{Chang} \& {Bildsten}}{{Chang} \&
  {Bildsten}}{2004}]{2004ApJ...605..830C}
{Chang} P.,  {Bildsten} L.,  2004, \mn@doi [\apj] {10.1086/382271}, \href
  {http://adsabs.harvard.edu/abs/2004ApJ...605..830C} {605, 830}

\bibitem[\protect\citeauthoryear{{Chang}, {Bildsten}  \& {Arras}}{{Chang}
  et~al.}{2010}]{2010ApJ...723..719C}
{Chang} P.,  {Bildsten} L.,   {Arras} P.,  2010, \mn@doi [\apj]
  {10.1088/0004-637X/723/1/719}, \href
  {http://adsabs.harvard.edu/abs/2010ApJ...723..719C} {723, 719}

\bibitem[\protect\citeauthoryear{{Cumming}, {Brown}, {Fattoyev}, {Horowitz},
  {Page}  \& {Reddy}}{{Cumming} et~al.}{2017}]{2017PhRvC..95b5806C}
{Cumming} A.,  {Brown} E.~F.,  {Fattoyev} F.~J.,  {Horowitz} C.~J.,  {Page} D.,
    {Reddy} S.,  2017, \mn@doi [\prc] {10.1103/PhysRevC.95.025806}, \href
  {https://ui.adsabs.harvard.edu/\#abs/2017PhRvC..95b5806C} {95, 025806}

\bibitem[\protect\citeauthoryear{{Degenaar} et~al.,}{{Degenaar}
  et~al.}{2014}]{2014ApJ...791...47D}
{Degenaar} N.,  et~al., 2014, \mn@doi [\apj] {10.1088/0004-637X/791/1/47},
  \href {https://ui.adsabs.harvard.edu/\#abs/2014ApJ...791...47D} {791, 47}

\bibitem[\protect\citeauthoryear{{Degenaar} et~al.,}{{Degenaar}
  et~al.}{2019}]{2019arXiv190707205D}
{Degenaar} N.,  et~al., 2019, arXiv e-prints, \href
  {https://ui.adsabs.harvard.edu/abs/2019arXiv190707205D} {p. arXiv:1907.07205}

\bibitem[\protect\citeauthoryear{{Deibel}, {Cumming}, {Brown}  \&
  {Page}}{{Deibel} et~al.}{2015}]{2015ApJ...809L..31D}
{Deibel} A.,  {Cumming} A.,  {Brown} E.~F.,   {Page} D.,  2015, \mn@doi [\apj]
  {10.1088/2041-8205/809/2/L31}, \href
  {https://ui.adsabs.harvard.edu/\#abs/2015ApJ...809L..31D} {809, L31}

\bibitem[\protect\citeauthoryear{{Gudmundsson}, {Pethick}  \&
  {Epstein}}{{Gudmundsson} et~al.}{1983}]{1983ApJ...272..286G}
{Gudmundsson} E.~H.,  {Pethick} C.~J.,   {Epstein} R.~I.,  1983, \mn@doi [\apj]
  {10.1086/161292}, \href {http://adsabs.harvard.edu/abs/1983ApJ...272..286G}
  {272, 286}

\bibitem[\protect\citeauthoryear{{Haensel} \& {Zdunik}}{{Haensel} \&
  {Zdunik}}{1990}]{1990A&A...227..431H}
{Haensel} P.,  {Zdunik} J.~L.,  1990, \aap, \href
  {https://ui.adsabs.harvard.edu/\#abs/1990A&A...227..431H} {227, 431}

\bibitem[\protect\citeauthoryear{{Homan}, {Fridriksson}, {Wijnands}, {Cackett},
  {Degenaar}, {Linares}, {Lin}  \& {Remillard}}{{Homan}
  et~al.}{2014}]{2014ApJ...795..131H}
{Homan} J.,  {Fridriksson} J.~K.,  {Wijnands} R.,  {Cackett} E.~M.,  {Degenaar}
  N.,  {Linares} M.,  {Lin} D.,   {Remillard} R.~A.,  2014, \mn@doi [\apj]
  {10.1088/0004-637X/795/2/131}, \href
  {https://ui.adsabs.harvard.edu/abs/2014ApJ...795..131H} {795, 131}

\bibitem[\protect\citeauthoryear{{Liu}, {Matsuo}, {Hashimoto}, {Noda}  \&
  {Fujimoto}}{{Liu} et~al.}{2017}]{2017JPSJ...86l3901L}
{Liu} H.,  {Matsuo} Y.,  {Hashimoto} M.-a.,  {Noda} T.,   {Fujimoto} M.~Y.,
  2017, \mn@doi [Journal of the Physical Society of Japan]
  {10.7566/JPSJ.86.123901}, \href
  {https://ui.adsabs.harvard.edu/\#abs/2017JPSJ...86l3901L} {86, 123901}

\bibitem[\protect\citeauthoryear{{Merritt} et~al.,}{{Merritt}
  et~al.}{2016}]{2016ApJ...833..186M}
{Merritt} R.~L.,  et~al., 2016, \mn@doi [\apj] {10.3847/1538-4357/833/2/186},
  \href {https://ui.adsabs.harvard.edu/\#abs/2016ApJ...833..186M} {833, 186}

\bibitem[\protect\citeauthoryear{{Ootes}, {Page}, {Wijnands}  \&
  {Degenaar}}{{Ootes} et~al.}{2016}]{2016MNRAS.461.4400O}
{Ootes} L.~S.,  {Page} D.,  {Wijnands} R.,   {Degenaar} N.,  2016, \mn@doi
  [\mnras] {10.1093/mnras/stw1799}, \href
  {https://ui.adsabs.harvard.edu/\#abs/2016MNRAS.461.4400O} {461, 4400}

\bibitem[\protect\citeauthoryear{{Ootes}, {Wijnands}, {Page}  \&
  {Degenaar}}{{Ootes} et~al.}{2018}]{2018MNRAS.477.2900O}
{Ootes} L.~S.,  {Wijnands} R.,  {Page} D.,   {Degenaar} N.,  2018, \mn@doi
  [\mnras] {10.1093/mnras/sty825}, \href
  {https://ui.adsabs.harvard.edu/abs/2018MNRAS.477.2900O} {477, 2900}

\bibitem[\protect\citeauthoryear{{Page}}{{Page}}{2016}]{2016ascl.soft09009P}
{Page} D.,  2016, {NSCool: Neutron star cooling code}, Astrophysics Source Code
  Library (\mn@eprint {ascl} {1609.009})

\bibitem[\protect\citeauthoryear{{Page} \& {Reddy}}{{Page} \&
  {Reddy}}{2013}]{2013PhRvL.111x1102P}
{Page} D.,  {Reddy} S.,  2013, \mn@doi [\prl] {10.1103/PhysRevLett.111.241102},
  \href {https://ui.adsabs.harvard.edu/\#abs/2013PhRvL.111x1102P} {111, 241102}

\bibitem[\protect\citeauthoryear{{Parikh} et~al.,}{{Parikh}
  et~al.}{2017}]{2017ApJ...851L..28P}
{Parikh} A.~S.,  et~al., 2017, \mn@doi [\apj] {10.3847/2041-8213/aa9e03}, \href
  {https://ui.adsabs.harvard.edu/\#abs/2017ApJ...851L..28P} {851, L28}

\bibitem[\protect\citeauthoryear{{Potekhin}, {Chabrier}  \&
  {Yakovlev}}{{Potekhin} et~al.}{1997}]{1997A&A...323..415P}
{Potekhin} A.~Y.,  {Chabrier} G.,   {Yakovlev} D.~G.,  1997, \aap, \href
  {https://ui.adsabs.harvard.edu/#abs/1997A&A...323..415P} {323, 415}

\bibitem[\protect\citeauthoryear{{Potekhin}, {Yakovlev}, {Chabrier}  \&
  {Gnedin}}{{Potekhin} et~al.}{2003}]{2003ApJ...594..404P}
{Potekhin} A.~Y.,  {Yakovlev} D.~G.,  {Chabrier} G.,   {Gnedin} O.~Y.,  2003,
  \mn@doi [\apj] {10.1086/376900}, \href
  {https://ui.adsabs.harvard.edu/#abs/2003ApJ...594..404P} {594, 404}

\bibitem[\protect\citeauthoryear{{Potekhin}, {Pons}  \& {Page}}{{Potekhin}
  et~al.}{2015}]{2015SSRv..191..239P}
{Potekhin} A.~Y.,  {Pons} J.~A.,   {Page} D.,  2015, \mn@doi [\ssr]
  {10.1007/s11214-015-0180-9}, \href
  {https://ui.adsabs.harvard.edu/#abs/2015SSRv..191..239P} {191, 239}

\bibitem[\protect\citeauthoryear{{Waterhouse}, {Degenaar}, {Wijnands}, {Brown},
  {Miller}, {Altamirano}  \& {Linares}}{{Waterhouse}
  et~al.}{2016}]{2016MNRAS.456.4001W}
{Waterhouse} A.~C.,  {Degenaar} N.,  {Wijnands} R.,  {Brown} E.~F.,  {Miller}
  J.~M.,  {Altamirano} D.,   {Linares} M.,  2016, \mn@doi [\mnras]
  {10.1093/mnras/stv2959}, \href
  {https://ui.adsabs.harvard.edu/abs/2016MNRAS.456.4001W} {456, 4001}

\bibitem[\protect\citeauthoryear{{Wijnands}, {Degenaar}  \& {Page}}{{Wijnands}
  et~al.}{2017}]{2017JApA...38...49W}
{Wijnands} R.,  {Degenaar} N.,   {Page} D.,  2017, \mn@doi [Journal of
  Astrophysics and Astronomy] {10.1007/s12036-017-9466-5}, \href
  {https://ui.adsabs.harvard.edu/#abs/2017JApA...38...49W} {38}

\bibitem[\protect\citeauthoryear{{Wijngaarden}, {Ho}, {Chang}, {Heinke},
  {Page}, {Beznogov}  \& {Patnaude}}{{Wijngaarden}
  et~al.}{2019}]{2019MNRAS.484..974W}
{Wijngaarden} M.~J.~P.,  {Ho} W. C.~G.,  {Chang} P.,  {Heinke} C.~O.,  {Page}
  D.,  {Beznogov} M.,   {Patnaude} D.~J.,  2019, \mn@doi [\mnras]
  {10.1093/mnras/stz042}, \href
  {https://ui.adsabs.harvard.edu/\#abs/2019MNRAS.484..974W} {484, 974}

\makeatother
\end{thebibliography}

% Alternatively you could enter them by hand, like this:
% This method is tedious and prone to error if you have lots of references
%\begin{thebibliography}{99}
%\bibitem[\protect\citeauthoryear{Author}{2012}]{Author2012}
%Author A.~N., 2013, Journal of Improbable Astronomy, 1, 1
%\bibitem[\protect\citeauthoryear{Others}{2013}]{Others2013}
%Others S., 2012, Journal of Interesting Stuff, 17, 198
%\end{thebibliography}

%%%%%%%%%%%%%%%%%%%%%%%%%%%%%%%%%%%%%%%%%%%%%%%%%%

%%%%%%%%%%%%%%%%% APPENDICES %%%%%%%%%%%%%%%%%%%%%

%\appendix

%%%%%%%%%%%%%%%%%%%%%%%%%%%%%%%%%%%%%%%%%%%%%%%%%%

% Don't change these lines
\bsp	% typesetting comment
\label{lastpage}
\end{document}